# Wear and corrosion properties of Mg(OH)$_2$ compound layer formed on magnesium alloy in superheated water vapor

Tianxiang Peng, Liang Wang[*]

Corresponding to: wlimt@dlmu.edu.cn

Department of Materials Science and Engineering, Dalian Maritime University, Dalian 116026, China

**Abstract:** The wear resistance of magnesium and its alloys is very poor due to their inherent low hardness and poor tribological properties. Improving the wear resistance of magnesium and its alloys plays a significant role in expanding their application range. Various surface treatments have been used to produce protective coatings or layers for solving this problem. In this work, the Mg(OH)$_2$ layer with a thickness about 25 μm was directly formed on AZ91 magnesium alloy by oxidizing in superheated water vapor. SEM and XRD were employed to characterize the microstructure and phase composition of oxidized layer. The effect of oxidized layer on the friction and wear properties of magnesium alloy against the bearing steel ball were evaluated by using a reciprocating friction and wear test apparatus. The results showed that layer exhibited low friction coefficients and extremely small wear loss compared to the untreated substrate.

## 1. Introduction

Magnesium and magnesium alloys are characterized by their exceptional lightness, making them some of the lightest structural materials available. They offer a high strength-to-weight ratio, which is a significant advantage in various industries, particularly in aerospace and automotive where weight reduction is crucial for fuel efficiency and performance. However, their inherent low hardness and poor tribological properties pose challenges for widespread adoption, particularly in applications involving sliding and wearing parts. In addition, their corrosion resistance is very poor in aggressive media or even in moisture environments. To enhance their corrosion and wear resistance, surface treatment technologies are necessary for modifying the surface of magnesium and its alloys to form a protective layer, which in turn improves their corrosion and wear resistance. Commonly used surface treatment methods include anodic oxidation, chemical conversion, plasma electrolytic oxidation, electroplating, friction stir processing, cladding, physical vapor deposition and others [1-10]. As these methods create a dense and adhesive protective film or coating on the surface of magnesium and its alloys, the corrosion and wear can be reduced or prevented by hindering the invasion of corrosive media and wear of counterpart. Zhao et al. reported that a corrosion inhibitors intercalated CI@LDH-MAO composite coating was fabricated on AZ31 Mg alloy by micro-arc oxidation and hydrothermal treatment. Under 5 N normal load, the coefficient of friction and wear rate were evidently reduced [11]. The wear resistance of AM50 Mg alloy was improved by forming a PEO coating with the inertly incorporated $SiO_2$ and $Si_3N_4$ particles [12]. Mao et al. prepared the composite coating on GW83 magnesium alloys for improving their wear resistance [13]. Zhang et al. deposited Fe-based amorphous coatings on the surface of WE43 magnesium alloy by the high velocity oxy-fuel (HVOF) spraying. The friction coefficient of the coating was only half that of the WE43 and the wear mechanism was changed [14]. The anodized coating produced from a phosphate solution on AZ91D magnesium alloy. The Wear loss of anodized magnesium alloy was greatly reduced in comparison with the substrate [15]. The tribological properties of AZ31 magnesium alloy were improved by depositing DLC/AlN/Al protective coating which were of high hardness, wear resistance and low friction coefficient [16]. To improve corrosion and wear resistance of AZ31magnesium alloy, Xie et al. deposited the Ti-DLC monolayer and Ti/Ti-DLC multilayer films using filter cathodic vacuum arc deposition technology [17].

The purpose of the present paper is to describe and analyze the wear properties of $Mg(OH)_2$ layer formed on magnesium alloy oxidized in superheated water vapor.

## 2. Experimental

The magnesium alloy used in this study is AZ91D, with a chemical composition (wt.%) of Al (8.3-9.7), Zn (0.35-1.0), Mn (0.15-0.5) and Mg (balance). Specimens with a thickness of 5 mm were cut from a bar with a diameter of 12 mm. Prior to treatment, the samples surfaces were mechanically abraded using silicon carbide papers (400-1200grit). Figure 1 illustrates the container utilized for processing the magnesium alloy. The vessel is constructed from stainless steel, with a total volume of 4.4 liters. The magnesium alloy samples were placed on a sample holder within the container. An appropriate amount of tap water was added to the container, ensuring that the all water could be converted into vapor at the selected heating temperature without excessively increasing the pressure inside the container, thus avoiding any potential hazards. The container was then sealed, heated to reach a required temperature of 200oC and held at this temperature for 4 h. Apart from the sandpaper abrasion, no any additional pretreatments were required for the samples. The process of layer formation consumed only electrical energy and a very little amount of water, with no other consumables or emissions.

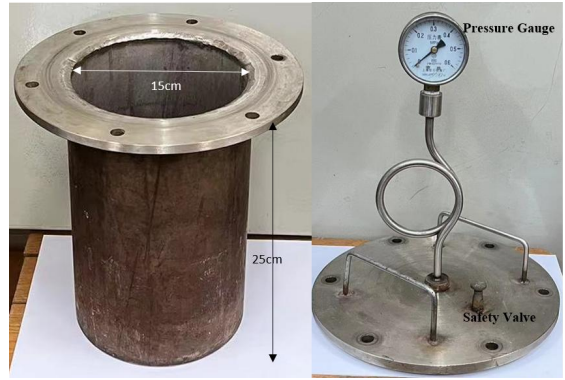

Fig.1 Photograph of the container used to prepare Mg(OH)2 on magnesium alloy.

The crystal structure of the samples was examined by X-ray diffraction (XRD) using Cu K radiation ($\lambda$=1.54 Å) on a Rigaku D/Max-Ultima using a conventional Bragg-Brentano configuration. The tube acceleration voltage and current used were 40 kV and 40 mA respectively. The scan range of the 2 theta was from 30 to 80 with an increment of 0.02°/step and a scan rate of 2°/min. SUPRA 55 scanning electron microscope with an EDX analysis system was used to observe the surface morphology and polished cross-sections of the samples (which were sputtered with a thin gold layer in order to prevent surface charging effects). Layer thickness was measured using microscopy on polished sections.

A HRS-2M tribometer with a pin-on-disc configuration was used to characterize the tribological behavior of untreated and oxidized magnesium alloy samples at room temperature with a relative humidity of 40% under dry sliding conditions. A GCr15 bearing steel ball with a diameter of 5 mm was used as the friction counter body. A normal load of 3 N, 5 N and 8 N was applied to the steel ball. Reciprocating length was 5 mm and reciprocating frequency was 5 Hz with duration of 30 min. The friction coefficient was continuously monitored during the test. After the wear tests, the morphology and section profile of the wear tracks were analyzed by SEM and confocal laser scanning microscopy (CLSM-Olympus OLS4000). The corrosion property of the layer was examined in 3.5% NaCl solution by immersion tests.

## 3. Results and discussion

Fig. 2 shows appearances of original and oxidized magnesium alloy samples and polished cross-sections of Mg(OH)$_2$ layer formed on AZ91.

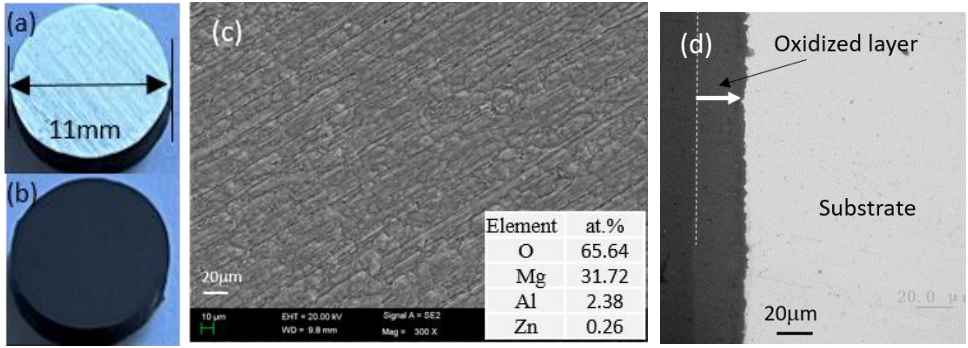

Fig. 2. Appearances of magnesium alloy samples: (a) original, (b) oxidized, (c) surface and (d) cross-section microstructure of Mg(OH)$_2$ layer.

From Fig. 2 (a) and (b), the appearance of untreated magnesium alloy sample, when viewed by the naked eye, displayed a typical metallic luster on their surface. However, following oxidation treatment, the surface assumed a uniform deep black coloration. Fig. 2 (c) shows the surface morphology of the oxidized layer by SEM and the composition provided by EDX analysis. It can be observed that the atomic ratio between oxygen and magnesium in the layer is very close to 2:1, confirming that a layer composed of magnesium hydroxide has been obtained after oxidation in water vapor following the reaction Mg + 2H$_2$O → Mg(OH)$_2$ + H$_2$. From Fig. 2 (d), it can be see that the oxidized layer was also black observed under the optical microscope. The layer was dense and homogeneous without evident defects. Therefore, it is very easy to measure the thickness of the layer. The thickness of the layer is about 25 μm. There was no any problem for adhesion between layer and matrix due to forming the layer through oxidization of magnesium.

Fig. 3 presents the XRD patterns of untreated and oxidized AZ91 magnesium alloy. The untreated AZ91 alloy mainly composed of hexagonal Mg phase. After oxidation treatment, the diffraction peak intensities corresponding to Mg all weakened relatively. Distinct diffraction peaks emerged at 32.8°, 38.1°, 58.7°, 62.2°, and 68.9°, which are attributed to the diffraction generated by the (100), (101), (110), (111), and (200) crystallographic planes of Mg(OH)$_2$, respectively. Despite the thickness of the layer reaching about 25 μm, the diffraction peaks originating from the substrate maintained a significant level of intensity. The reason is that the constituent elements magnesium and oxygen in the layer have low mass absorption coefficients for X-rays, especially oxygen, which has a mass absorption coefficient for Cu radiation of less than one-third that of copper. Therefore, diffraction peaks originating from the substrate still retain considerable intensity.

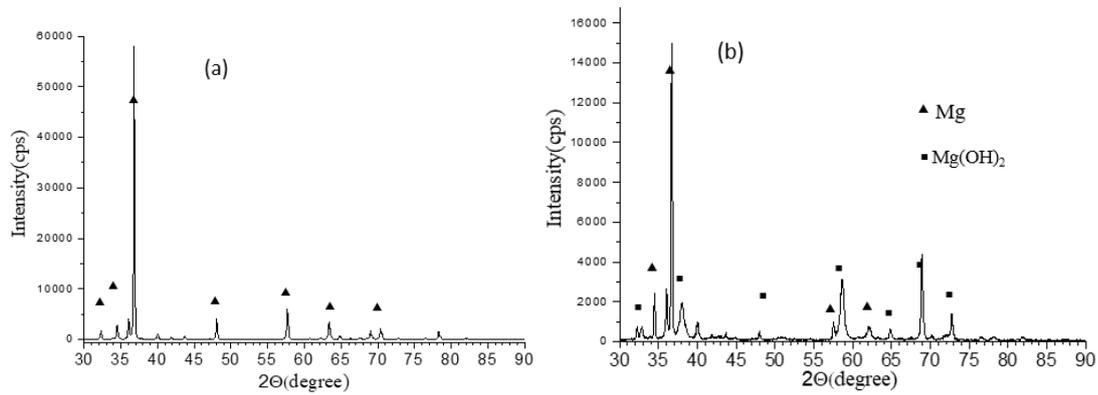

Fig. 3. XRD analysis of AZ91 alloy samples: (a) untreated, (b) treated in 200°C superheated water vapor for 2 h.

The wear scars of untreated sample after sliding 30 min under a load of 3, 5 and 10 N were observed by SEM and illustrated in Fig. 4.

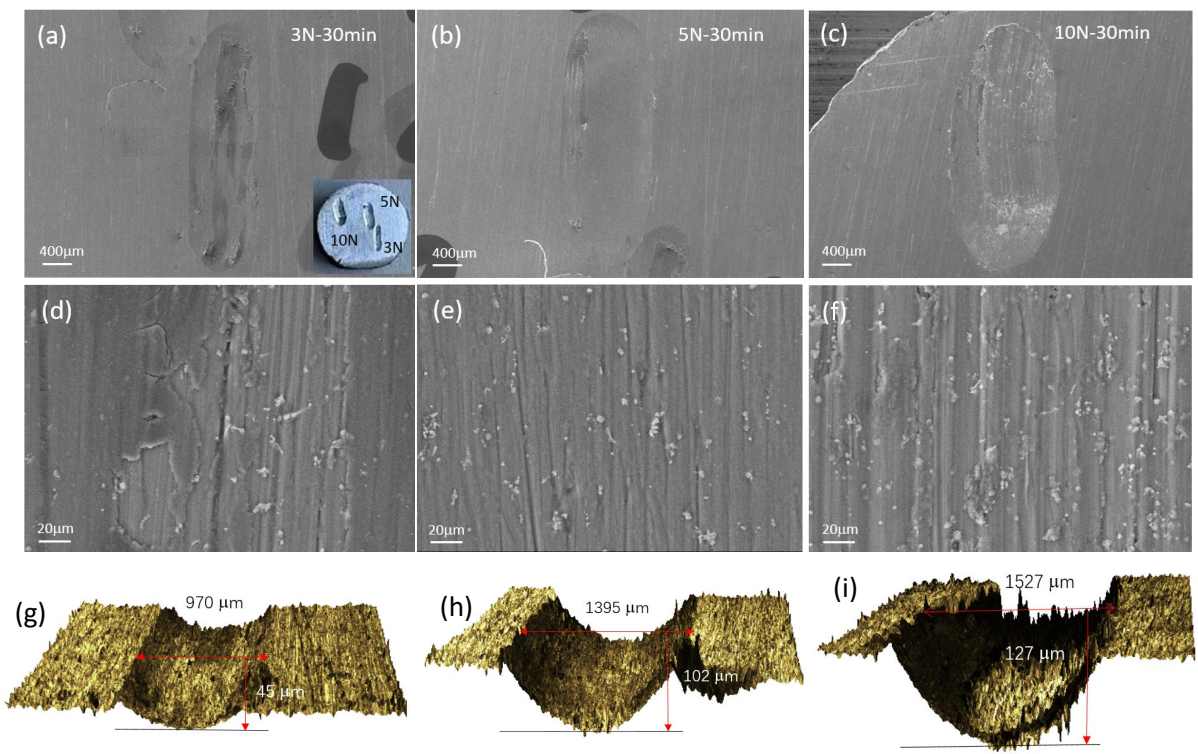

Fig. 4. The micrograph of wear tracks of the untreated AZ91 magnesium alloy under different loads: (a, d, g) for 3 N, (b, e, h) for 5 N and (c, f, i) for 10 N.

SEM and 3D morphologies of wear track with different magnifications after sliding 30 min under 3 N load are displayed in Fig. 4 (a, d, g). It is observed that after wear test, the resultant wear track on the sample surface measured approximately 930~1000 μm in width. A detailed examination revealed that the evident oxide adhesion and scratches were visible in the enlarged wear track areas. The depth of wear scar reached about 45 μm. At a load of 5N, a marked increase in the width of the wear track is observed, reaching 1320~1400 μm with the wear tracks predominantly characterized by grooves and minimal oxide deposits. The depth of wear scar increased to about 102 μm. With the load increasing to 10 N, the wear track generated is distinctly different from those produced by 3N and 5N loads. In addition to the further increase in wear track width to 1520~1600 μm, the worn region exhibited severe plastic deformation, as well as signs of adhesion and tearing accompanied by with evident furrows. The wear scar had a depth reaching about 127 μm. These results indicate that the load had a significant impact on the tribological properties of magnesium alloys. At lower load conditions, the dominant wear mechanisms are oxidation and ploughing. However, with the increase in load, a combination of ploughing, plastic deformation and adhesion was the wear mechanism.

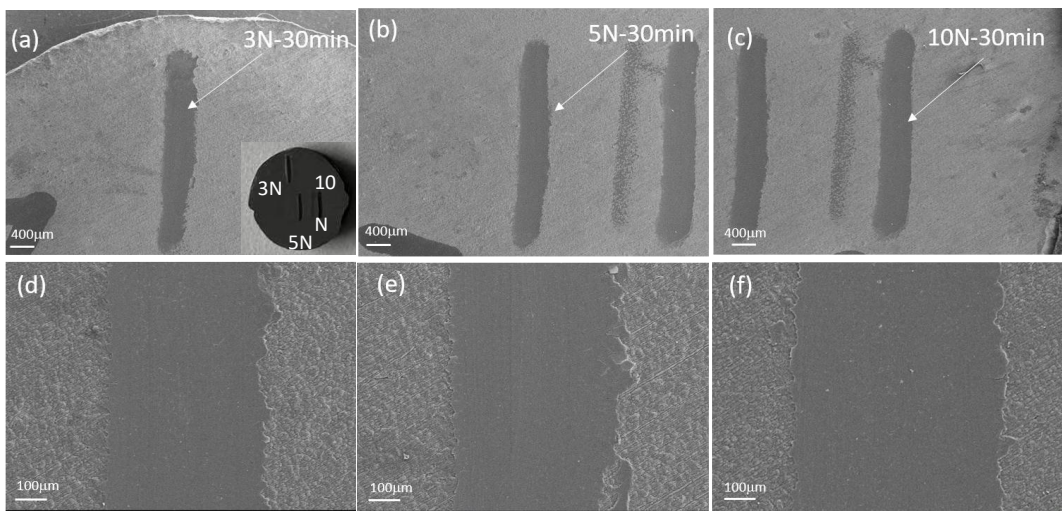

Fig. 5. SEM images of wear scar on treated AZ91 magnesium alloy after test under three loads.

The SEM morphologies of wear tracks created on the oxidized sample surface under three different loads are presented in Fig. 5. Compared to the untreated samples, under the same wear conditions, there is no obvious difference in the morphology of the wear tracks obtained for three loads. The width of the wear tracks ranges from about 500 to 600 μm under the three loads after sliding for 30 min. The wear track areas are very smooth and level, with no observable scratches, resembling a polished appearance. After 30 minutes of wear, in addition to the clearly visible wear tracks on the sample surface, no signs of scratches, adhesion, or plastic deformation were observed. The wear mechanism appears to be in the form of grinding and polishing. The results show that significant improvements in the wear resistance of AZ91 magnesium alloy can be achieved by treatment in superheated water vapor.

3 and 2 D profilometry of wear tracks left on the oxidized sample are presented in Fig. 6. The width of the wear tracks left on the surface of the oxidized samples was significantly reduced compared to the untreated samples, measuring 535, 562 and 622 μm for 3N, 5N and 10 N load, respectively. Additionally, the depth of the wear tracks was also significantly reduced to about 5, 7 and 12 μm for 3, 5 and 10 N load respectively. The narrow and shallow wear track of oxidized sample also indicated that the wear loss was dramatically reduced.

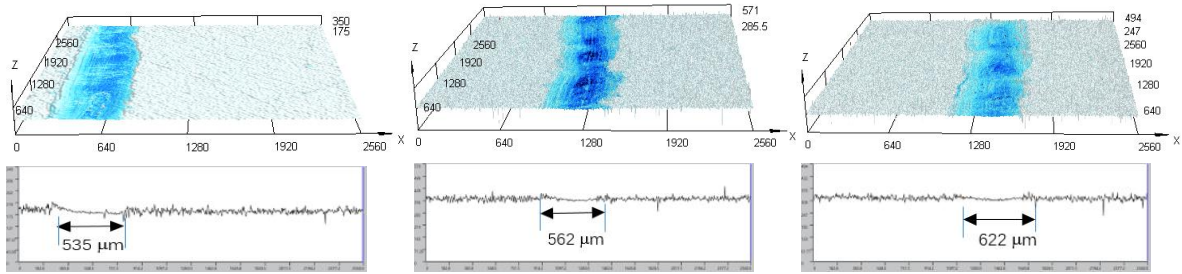

Fig. 6. Three and two dimensional profilometry of wear tracks left on the oxidized sample after sliding wear for 30 min at 3 N, 5 N and 10 N respectively.

The friction coefficient and wear rate for untreated and treated samples underwent sliding for 30 min under a load of 3, 5 and 10 N are shown in Fig. 7.

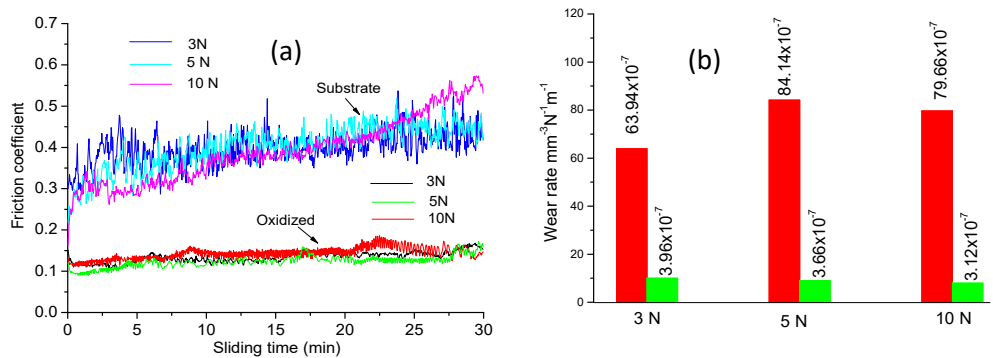

Fig. 7. The curves of friction coefficient (a) and wear rate (b) of untreated and treated AZ91 magnesium alloy after sliding for 30 min under different loads.

For the untreated magnesium alloy sample, the friction coefficient quickly rose to about 0.4 once the test started, then fluctuated between 0.35 and 0.45 under 3 N and 5 N loads. In contrast, as the load increased to 10 N, the friction coefficient displayed a progressive increase until the end of the test, rising from an initial value of around 0.3 to about 0.55, albeit with a slightly diminished amplitude of fluctuation. Compared to the untreated sample, the friction coefficients of the oxidized sample significantly decreased under the action of all three loads. The coefficients remained predominantly within the range of 0.12 to 0.17. The fluctuation was also evidently decreased within a very narrow scope. These results suggested that the magnesium hydroxide layer formed on the surface of magnesium alloy by oxidation possessed superior tribological characteristics. The wear rate of untreated sample was 63.95 ×$10^{-7}$ $mm^3N^{-1}.m^{-1}$ for 3 N, 84.14×$10^{-7}$ $mm^3N^{-1}.m^{-1}$ for 5 N and 79.66×$10^{-7}$ $mm^3N^{-1}.m^{-1}$ for 10 N. For oxidized sample, the wear rate was 3.96×$10^{-7}$ $mm^3N^{-1}.m^{-1}$ for 3 N, 3.62×$10^{-7}$ $mm^3N^{-1}.m^{-1}$ for 5N and 3.12×$10^{-7}$ $mm^3N^{-1}.m^{-1}$ for 10 N. Compared to the untreated samples, the wear rates are reduced by approximately 16, 23, and 25 times, respectively. Low friction coefficient and wear rate meant that the tribological properties of magnesium alloy were greatly improved by forming a Mg(OH)2 layer in water vapor. The wear resistance of magnesium alloy was improved by depositing Ti-DLC monolayer and Ti/Ti-DLC multilayer films with filter cathodic vacuum arc deposition technology. Under a load of 5 N and a frequency of 1 HZ, the friction coefficient of all coatings was below 0.12, the wear rate was 2–3 orders of magnitude lower than that of uncoated the magnesium alloy [17]. Das et al., reported a high wear resistant TiC/Co/Y2O3 composite coating which was synthesized on the surface of AZ91D Mg alloy by plasma transferred arc coating process. The wear rate was reduced about seven times as compared to AZ91D Mg [18]. Chen et al. used ultrasonic impact treatment to simultaneously fabricate surface Ti-MoS2 self-lubricating coatings on AZ91Dmagnesium alloy. The coated samples exhibited superior anti-wear properties. Compared to untreated samples, the wear volume at 5 N, 15 N, and 25 N loads was reduced by 89.78 %, 84.23 %, and 43.88 %, respectively [8].To achieve superior corrosion and wear resistance, Wang et al. prepared a Ni–WC/Al–Ni functionally graded coating on AZ91D Mg alloy. The wear resistance of the coating was 5–8 times that of the substrate [19]. Ding et al. produced a novel Nb2O5/Mg gradient coating on AZ31 magnesium alloy substrate by using a sputtering deposition process. The wear rate was reduced by more than an order of magnitude [20]. DLC films show a stable and relatively low friction coefficient under dry conditions at room temperature, in comparison with uncoated AZ91magnesium alloy [21]. Aluminum matrix composite coatings reinforced with SiCp were deposited using oxyacetylene thermal spraying on ZE41 magnesium alloy substrates. The incorporation of SiCp to the spraying mixture increased wear resistance up to 2 times more than the initial ZE41 substrates [22].

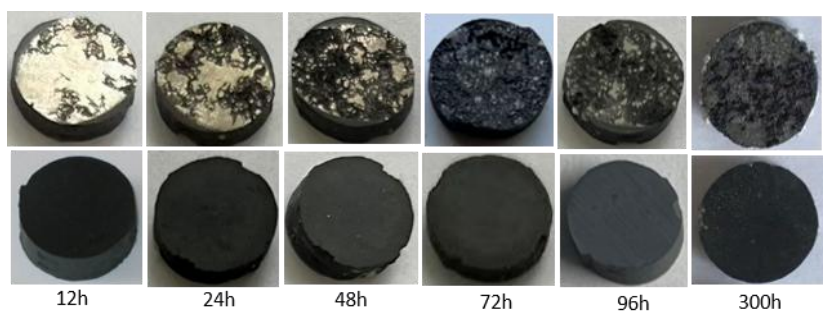

Fig. 8. Appearances of untreated and oxidized magnesium alloy immersed in a 3.5% NaCl solution for various times.

Corrosive immersion test was carried out to evaluate corrosion resistance of untreated and treated samples. Surface morphologies of two samples after being immersed in a 3.5% NaCl solution for different times are shown in Fig. 8. As illustrated, following a 12 h immersion period, localized areas on the sample surface exhibited signs of corrosion, with distinct corrosion byproducts becoming evident upon examination. As the duration of immersion increased, the area of corrosion progressively expanded, and the quantity of corrosion products continuously accumulated. After 48 h of immersion, a significant portion of the sample surface was covered by corrosion products. Upon reaching 72 h of immersion, the entire sample surface exhibited severe corrosion. After 300 h of immersion, in addition to the surface, parts of the sample had completely corroded away. In contrast, the surface of the oxidized sample exhibited negligible alterations over the duration of the immersion period. Even after 300 h of immersion, the sample surface remained largely unchanged, and no corrosion products were observed on the surface. Clearly, the magnesium hydroxide film layer on the surface of magnesium alloys can effectively enhance their corrosion resistance through mitigating the onset and progression of corrosion processes.

## 4. Conclusions

$Mg(OH)_2$ layer with a thickness about 25 μm was successfully formed on magnesium alloy after oxidized in superheated water vapor atmosphere at 200 °C for 4 h. A reciprocating friction and wear testing machine was used to conduct comparative tribological tests on untreated samples and oxidized samples under 3N, 5N, and 10N loads with dry friction conditions. The formation of a magnesium hydroxide layer significantly reduced the friction coefficient of the magnesium alloy, from 0.4-0.5 for the untreated samples to 0.12-0.17. The wear rates were also reduced from 63.95 ×$10^{-7}$ $mm^3N^{-1}.m^{-1}$ to 3.96×$10^{-7}$ $mm^3N^{-1}.m^{-1}$ for 3N, 84.14×$10^{-7}$ $mm^3N^{-1}.m^{-1}$ to 3.62×$10^{-7}$ $mm^3N^{-1}.m^{-1}$ for 5N, and 79.66×$10^{-7}$ $mm^3N^{-1}.m^{-1}$ to 3.12×$10^{-7}$ $mm^3N^{-1}.m^{-1}$ for 10 N. In addition, the corrosion resistance of the untreated and oxidized magnesium alloy was evaluated using immersion test in a 3.5 wt.% NaCl solution. The corrosion resistance of magnesium alloy also significantly improved due to the formation of $Mg(OH)_2$ layer.